\documentclass[12pt]{article}
\pdfoutput=1

\usepackage{amsmath}
\usepackage{amsfonts}
\usepackage{amscd}
\usepackage{amsthm}
\usepackage{setspace}

\usepackage{graphicx}
\usepackage{authblk}
\usepackage{caption}
\usepackage{ytableau}
\usepackage{mathtools}

\def\Z{\mathbb{Z}}

\def\C{\mathbb{C}}
\def\P{\mathbb{P}}

\def\til{\tilde}

\begin{document}

\begin{titlepage}

\begin{flushright}
KEK-TH 2036
\end{flushright}

\vskip 1cm

\begin{center}

{\large F-theory models on K3 surfaces with various Mordell-Weil ranks \\ 
-constructions that use quadratic base change of rational elliptic surfaces}

\vskip 1.2cm

Yusuke Kimura$^1$ 
\vskip 0.4cm
{\it $^1$KEK Theory Center, Institute of Particle and Nuclear Studies, KEK, \\ 1-1 Oho, Tsukuba, Ibaraki 305-0801, Japan}
\vskip 0.4cm
E-mail: kimurayu@post.kek.jp

\vskip 1.5cm
\abstract{We constructed several families of elliptic K3 surfaces with Mordell--Weil groups of ranks from 1 to 4. We studied F-theory compactifications on these elliptic K3 surfaces times a K3 surface. 
\par Gluing pairs of identical rational elliptic surfaces with nonzero Mordell--Weil ranks yields elliptic K3 surfaces, the Mordell--Weil groups of which have nonzero ranks. The sum of the ranks of the singularity type and the Mordell--Weil group of any rational elliptic surface with a global section is 8. By utilizing this property, families of rational elliptic surfaces with various nonzero Mordell--Weil ranks can be obtained by choosing appropriate singularity types. Gluing pairs of these rational elliptic surfaces yields families of elliptic K3 surfaces with various nonzero Mordell--Weil ranks. 
\par We also determined the global structures of the gauge groups that arise in F-theory compactifications on the resulting K3 surfaces times a K3 surface. $U(1)$ gauge fields arise in these compactifications.}  

\end{center}
\end{titlepage}

\tableofcontents
\section{Introduction}
F-theory \cite{Vaf, MV1, MV2} provides a useful framework for model building in particle physics. $SU(5)$ grand unified theories with matter fields in $SO(10)$ spinor representations are naturally realized in F-theory. The F-theory approach can also avoid the issue concerning weakly coupled heterotic strings addressed in \cite{Witten}. Local F-theory models \cite{DWmodel, BHV1, BHV2, DWGUT} have been mainly considered in recent progress on model building in F-theory. Global structures of F-theory models, however, must be studied to address the problems of gravity and the early universe.
\par In this study, we constructed several families of elliptic K3 surfaces with a section that have the Mordell--Weil groups of ranks from 1 to 4. We give the Weierstrass equations with parameters to describe these families of elliptic K3 surfaces with nonzero Mordell--Weil ranks.
\par In F-theory compactification, the Mordell--Weil rank of elliptic fibration of the compactification space is equal to the number of $U(1)$-gauge fields that arise in the resulting theory. Therefore, a $U(1)$-gauge field arises in F-theory compactifications on the families of K3 surfaces with nonzero Mordell--Weil ranks that we constructed in this study. Particularly, $U(1)^n$ arises in F-theory compactification on a K3 surface with the Mordell--Weil group of rank $n$ times a K3 surface. F-theory compactifications on elliptic fibrations with a global section have been discussed, for example, in \cite{MorrisonPark, MPW, BGK, BMPWsection, CKP, BGK1306, CGKP, CKPS, AL, LSW, CKPT, CGKPS, KimuraMizoguchi}. The presence of $U(1)$ symmetry in F-theory compactifications relates to the realization of GUT, because $U(1)$ symmetry in F-theory helps to explain characteristic properties of GUT, including the suppression of proton decay, and the mass hierarchies of leptons and quarks.
\par It is technically considerably difficult to construct an elliptic K3 surface with a given Mordell--Weil rank by manipulating the coefficients of the Weierstrass equation that gives an elliptic K3 surface. Furthermore, the Picard numbers of elliptic K3 surfaces vary depending on the complex structures. Owing to this nature of a K3 surface, fixing the singularity type of an elliptic K3 surface does not determine the Mordell--Weil rank.
\par In contrast, the sum of the ranks of the Mordell--Weil group and the singularity type of a rational elliptic surface with a global section is always 8. Thus, it is much easier, when compared to an elliptic K3 surface, to construct a rational elliptic surface with a given Mordell--Weil rank, by choosing an appropriate singularity type. Furthermore, gluing two identical rational elliptic surfaces with a global section yields an elliptic K3 surface with a section, as discussed in \cite{KRES} in the context of F-theory compactifications and the stable degeneration limit \cite{FMW, AM} \footnote{Recent progress on stable degenerations in F-theory/heterotic duality can be found, for example, in \cite{AHK, BKW, CGKPS, KRES}.}. The Mordell--Weil rank of the resulting K3 surface is generically identical to the original rational elliptic surface under this operation. By utilizing the properties that we mentioned above, in this study we construct families of elliptic K3 surfaces with the Mordell--Weil ranks from 1 to 4, by constructing families of rational elliptic surfaces with the Mordell--Weil groups of ranks from 1 to 4, and gluing pairs of isomorphic rational elliptic surfaces that belong to these families.
\par The process of gluing two isomorphic rational elliptic surfaces with a section to give an elliptic K3 surface is described by the quadratic base change of a rational elliptic surface \cite{KRES}. This particularly means that the substitution of a quadratic equation into the coordinate variable of the base $\P^1$ in the Weierstrass equation of a rational elliptic surface yields the Weierstrass equation of the resulting K3 surface, which is obtained as the quadratic base change of the rational elliptic surface. Gluing two identical rational elliptic surfaces together can be seen as the reverse of stable degeneration, in which a K3 surface splits into two rational elliptic surfaces.  
\par The structures of rational elliptic surfaces \footnote{There are rational elliptic surfaces lacking a global section. Halphen surfaces are examples of such surfaces. See, for example, \cite{DolgachevZhang, CantatDolgachev, Kimura1801} for discussions of the structures of Halphen surfaces. In this study, we considered only rational elliptic surfaces that have a global section.} have been studied in the literature. The singularity types and the corresponding Mordell--Weil groups of rational elliptic surfaces are classified in \cite{OS}. Pencils of rational elliptic surfaces that have the singularity types of rank 8, namely, extremal rational elliptic surfaces \footnote{\cite{MP} classified the types of singular fibers of the extremal rational elliptic surfaces.}, are deduced in \cite{Nar}. Pencils of rational elliptic surfaces with the singularity types of rank 7 are obtained in a recent study \cite{Kurumadani}. 
\par The Picard number of any rational elliptic surface with a section is 10. The ranks of the Mordell--Weil group and the singularity type of a rational elliptic surface with a global section are complementary, and the sum of the two ranks is always 8. Therefore, it follows from this fact that rational elliptic surfaces with a section, the ranks of the singularity types of which are strictly less than 8, in fact have Mordell--Weil ranks greater than or equal to 1. In this study, we constructed families of rational elliptic surfaces with specific singularity types of ranks less than 8. The quadratic base change of these families of rational elliptic surfaces yields families of elliptic K3 surfaces with nonzero Mordell--Weil ranks. The fact that the resulting elliptic K3 surfaces obtained as the quadratic base change of rational elliptic surfaces with nonzero Mordell--Weil ranks have the Mordell--Weil groups of positive ranks follows from the following relation that holds between the Mordell--Weil group of a rational elliptic surface and that of the resulting K3 surface: Because a global section of a rational elliptic surface $X$ lifts to a global section of K3 surface $S$ through the quadratic base change, the Mordell--Weil group of a rational elliptic surface $X$ is a subgroup of the Mordell--Weil group of K3 surface $S$ obtained as the quadratic base change of the rational elliptic surface $X$. Thus, the Mordell--Weil rank of the resulting K3 surface $S$ is always larger than or equal to the Mordell--Weil rank of rational elliptic surface $X$:
\begin{equation}
{\rm rk}\, {\rm MW} (S) \ge {\rm rk}\, {\rm MW} (X).
\end{equation}
Therefore, the quadratic base change of rational elliptic surfaces with nonzero Mordell--Weil ranks yields elliptic K3 surfaces with various nonzero Mordell--Weil ranks. The Mordell--Weil rank of the resulting K3 surface $S$ and that of rational elliptic surface $X$ are equal for a generic quadratic base change; the Mordell--Weil rank of K3 surface $S$ becomes strictly larger than the Mordell--Weil rank of the original rational elliptic surface $X$ when the parameters of the quadratic base change assume special values \footnote{\cite{KimuraMizoguchi} discusses some situations in which the Mordell--Weil ranks of K3 surfaces, obtained as the quadratic base changes of an extremal rational elliptic surface, are enhanced for specific quadratic base changes.}. We discuss these in detail in Section \ref{sec2}. In this study, we focused on the generic situations in which the Mordell--Weil rank remains unchanged through the base change. 
\par Concretely, in this study we constructed families of rational elliptic surfaces with a global section, that have the singularity types $E_7$, $D_7$, $E_6 A_1$, $E_6$, $D_6$, $D_5$, and $A_4$. These surfaces are familiar, and they have been well studied in the context of heterotic/F-theory duality \cite{BIKMSV, FMW, MT}. These rational elliptic surfaces have singularity types of ranks from 4 to 7, therefore, these rational elliptic surfaces have the Mordell--Weil groups of ranks from 1 to 4; the quadratic base change of these families of rational elliptic surfaces yields families of elliptic K3 surfaces with Mordell--Weil ranks from 1 to 4. 
\par In F-theory compactification, non-Abelian gauge symmetries that arise on the 7-branes correspond to the types of singular fibers of elliptic fibration of the compactification space \cite{MV2, BIKMSV}. The correspondences of the types of singularities of the compactification space and the types of the singular fibers are listed in Table \ref{tabsingularity}. The singular fiber types of elliptic surfaces were classified in \cite{Kod1, Kod2}. See \cite{Ner, Tate} for discussions on determining the singular fibers of elliptic surfaces. Discussions on elliptic surfaces and singular fibers can also be found in \cite{Shiodamodular, Shioda, Silv, BHPV, SchShio}. 

\begingroup
\renewcommand{\arraystretch}{1.1}
\begin{table}[htb]
\begin{center}
  \begin{tabular}{|c|c|} \hline
$\begin{array}{c}
\mbox{Type of} \\
\mbox{singular fiber}
\end{array}$ & Singularity type \\ \hline
$I_1$ & none. \\
$II$ & none. \\
$I_n$ ($n\ge 2$) & $A_{n-1}$ \\
$I^*_m$ ($m\ge 0$) & $D_{4+m}$ \\
$III$ & $A_1$ \\
$IV$ & $A_2$ \\
$II^*$ & $E_8$ \\
$III^*$ & $E_7$ \\
$IV^*$ & $E_6$ \\ \hline   
\end{tabular}
\caption{Correspondences of types of singular fibers and the singularity types of the compactification space.}
\label{tabsingularity}
\end{center}
\end{table}  
\endgroup 

\par This note is structured as follows: The general strategy to construct families of elliptic K3 surfaces with the nonzero Mordell--Weil ranks from families of rational elliptic surfaces is described in Section \ref{sec2}. We describe the construction of rational elliptic surfaces with a section with specific singularity types in Section \ref{sec3}. As described in Section \ref{sec4}, after a brief review of the quadratic base change of a rational elliptic surface, we obtained several families of elliptic K3 surfaces with nonzero Mordell--Weil ranks as the quadratic base change of rational elliptic surfaces, that were constructed as described in Section \ref{sec3}. The resulting K3 surfaces have the Mordell--Weil groups of ranks from 1 to 4. The gauge groups arising in F-theory compactifications on these elliptic K3 surfaces are discussed in Section \ref{sec5}. Central results of this note are deduced in Section \ref{sec4} and Section \ref{sec5}. We state the concluding remarks in Section \ref{sec6}.

\section{General strategy to construct elliptic K3 surfaces with various Mordell--Weil ranks}
\label{sec2}
We describe the general strategy for constructing families of elliptic K3 surfaces with nonzero Mordell--Weil ranks in this section. 
\par We consider the gluing of two identical rational elliptic surfaces with a global section to obtain such K3 surfaces. Gluing pairs of rational elliptic surfaces with a section and deformations to K3 surfaces, as the reverse of the stable degeneration of a K3 surface splitting into a pair of rational elliptic surfaces, is discussed in \cite{KRES}. Gluing two identical rational elliptic surfaces along a pair of isomorphic smooth fibers corresponds to the quadratic base change of the rational elliptic surface \cite{KRES}. Because a pair of identical rational elliptic surfaces are glued together, the singular fibers of the resulting K3 surface are twice the number of the singular fibers of the rational elliptic surface. 
\par We show that a section of the rational elliptic surface $X$ extends to a global section of the elliptic K3 surface $S$ that is obtained as the quadratic base change of the rational elliptic surface $X$ through the quadratic base change. This can be shown as follows: We use $[t:s]$ to denote the coordinate of the base $\P^1$. The quadratic base change is an operation that replaces the coordinate variables $t,s$ of the base $\P^1$ with homogeneous quadratic polynomials $q_1(t,s), q_2(t,s)$:
\begin{eqnarray}
\label{base change in 2}
t  & \rightarrow & q_1(t,s) \\ \nonumber
s  & \rightarrow & q_2(t,s).
\end{eqnarray}
When $[h_1(t,s): h_2(t,s):h_3(t,s)]$ gives a section of the Weierstrass form of a rational elliptic surface $X$, where $h_i, i=1,2,3$, are functions over the base $\P^1$, $[h_1(q_1,q_2): h_2(q_1,q_2):h_3(q_1,q_2)]$ gives a section of the resulting elliptic K3 surface $S$ obtained as the quadratic base change. This shows that a section of a rational elliptic surface $X$ extends globally to a section of the resulting elliptic K3 surface $S$ through the quadratic base change. Therefore, the quadratic base change (\ref{base change in 2}) induces an injective \footnote{The quadratic base change is an operation in which a quadratic equation is substituted into the coordinate variable of the base $\P^1$; therefore, only the constant zero section of the rational elliptic surface $X$ is mapped to the zero section of the resulting elliptic K3 surface $S$. Thus, the induced group homomorphism is injective.} group homomorphism from the Mordell--Weil group MW$(X)$ of the rational elliptic surface $X$ to the Mordell--Weil group MW$(S)$ of the resulting elliptic K3 surface $S$. The Mordell--Weil group MW$(X)$ of the rational elliptic surface $X$ embeds into the Mordell--Weil group MW$(S)$ of the resulting elliptic K3 surface $S$ under the group homomorphism. Therefore, the Mordell--Weil group of the rational elliptic surface $X$ is a subgroup of the Mordell--Weil group of the resulting elliptic K3 surface $S$. When the Mordell--Weil group of the rational elliptic surface $X$ has rank $n$ and the sections, $s_1(t,s)$, $s_2(t,s)$, $\cdots$, $s_n(t,s)$, generate the free part $\Z^n$ of the Mordell--Weil group of the rational elliptic surface $X$, the sections, $s_1(q_1,q_2)$, $s_2(q_1,q_2)$, $\cdots$, $s_n(q_1,q_2)$, obtained through the quadratic base change generate a $\Z^n$ group in the Mordell--Weil group of the elliptic K3 surface $S$. 
\par Thus, it follows that the rank of the Mordell--Weil group MW$(S)$ of the resulting elliptic K3 surface $S$ is larger than or equal to the rank of the Mordell--Weil group MW$(X)$ of the original rational elliptic surface $X$. We show that the Mordell--Weil rank of the resulting elliptic K3 surface $S$ is equal to that of the original rational elliptic surface $X$ for generic values of the parameters of the quadratic base change. This can be proved as follows: Suppose that the Mordell--Weil rank of the K3 surface $S$ is strictly larger than the Mordell--Weil rank of the original rational elliptic surface $X$ for generic values of the parameters of the quadratic base change. This means that the Mordell--Weil group MW$(S)$ of the K3 surface $S$ has a free section, $s_*$, that does not come from the sections of the original rational elliptic surface $X$ under the group homomorphism. As discussed in \cite{KRES}, the resulting elliptic K3 surface $S$ splits into a pair of two isomorphic original rational elliptic surfaces $X$ under the stable degeneration. Because the Mordell--Weil rank of the K3 surface $S$ is strictly larger than that of the rational elliptic surface $X$ for generic values of the parameters of the quadratic base change, the free section $s_*$ remains a free section under the stable degeneration; therefore, the free section $s_*$ splits into sections of two copies of rational elliptic surfaces $X$. We denote the resulting section of the rational elliptic surface $X$ by $\til{s}_*$. This means that the free section $s_*$ of the K3 surface $S$ can be obtained by gluing together these sections $\til{s}_*$ of the rational elliptic surfaces $X$. It follows that the section $\til{s}_*$ of the original rational elliptic surface $X$ is mapped to the free section $s_*$ of the elliptic K3 surface $S$ under the group homomorphism, which is a contradiction. Thus, we conclude that the rank of the Mordell--Weil group of the elliptic K3 surface $S$ is equal to the rank of the Mordell--Weil group of the original rational elliptic surface $X$ for generic values of the parameters of the quadratic base change.
\par The Picard number of an elliptic K3 surface with a global section depends on the complex structure, ranging from 2 to 20, and it varies; even when we fix the singularity type of an elliptic K3 surface, this does not determine the rank of the Mordell--Weil group. In contrast, the Picard number of any rational elliptic surface with a section is 10. It follows from the Shioda--Tate formula \cite{Shiodamodular, Tate1, Tate2} that the sum of the rank of the singularity type of a rational elliptic surface with a global section and the rank of the Mordell--Weil group is 8:
\begin{equation}
\label{sum ranks in 2}
{\rm rk}\, ADE + {\rm rk}\, {\rm MW} =8.
\end{equation}
We have used rk $ADE$ in (\ref{sum ranks in 2}) to denote the rank of the singularity type of a rational elliptic surface. When the rank of the singularity type of a rational elliptic surface is fixed, this property determines the Mordell--Weil rank. Using this fact, we construct families of rational elliptic surfaces with nonzero Mordell--Weil ranks, by choosing the singularity types of ranks less than 8. The quadratic base change of these rational elliptic surfaces with nonzero Mordell--Weil groups yields families of elliptic K3 surfaces with nonzero Mordell--Weil ranks. 
\par To be explicit, we constructed families of rational elliptic surfaces with Mordell--Weil ranks 1, 2, 3, and 4; the quadratic base change of these families generically yields families of elliptic K3 surfaces with the Mordell--Weil ranks 1, 2, 3, and 4. Families of such rational elliptic surfaces are described in Section \ref{sec3}, and families of elliptic K3 surfaces obtained as the quadratic base change are given in Section \ref{sec4}. As discussed in Section \ref{sec5}, $U(1)^n$ gauge groups, $n=1, \cdots, 4$, arise in F-theory compactifications on these K3 surfaces times a K3. 
\par As families of rational elliptic surfaces with the Mordell--Weil groups of rank 1, in this study, we considered rational elliptic surfaces with the following rank 7 singularity types:
\begin{equation}
\label{singularity rank 7 in 2}
E_7, \hspace{1mm} D_7, \hspace{1mm} E_6A_1.
\end{equation}
\par We considered rational elliptic surfaces with the singularity types:
\begin{equation}
E_6, \hspace{1mm} D_6
\end{equation}
to obtain families of rational elliptic surfaces with the Mordell--Weil rank 2. 
\par Rational elliptic surfaces with the singularity type
\begin{equation}
D_5
\end{equation}
have the Mordell--Weil rank 3.
\par We considered rational elliptic surfaces with the singularity type
\begin{equation}
\label{singularity rank 4 in 2}
A_4
\end{equation}
to obtain a family of rational elliptic surfaces with the Mordell--Weil rank 4.
\par The aforementioned argument applies to general rational elliptic surfaces. Namely, gluing two isomorphic rational elliptic surfaces with the Mordell--Weil rank $n$ along isomorphic smooth fibers generically yields an elliptic K3 surface with the Mordell--Weil rank $n$, the singular fibers of which are twice the number of the singular fibers of the original rational elliptic surface. $U(1)^n$ gauge symmetry arises in F-theory compactification on the resulting K3 surface times a K3 surface. For example, when the most general homogeneous polynomials of degree 4, $f$, and degree 6, $g$, are chosen, the Weierstrass equation
\begin{equation}
y^2=x^3+f\, x+g
\end{equation}
describes a rational elliptic surface with the singularity type of rank 0. By the Shioda--Tate formula, rational elliptic surfaces with the singularity types of rank 0 have the Mordell--Weil rank 8. The complex structure moduli of such rational elliptic surfaces is eight-dimensional. The Mordell--Weil groups of generic elliptic K3 surfaces obtained as the quadratic base change of rational elliptic surfaces with the Mordell--Weil rank 8 generically have rank 8. 
\par Applying the technique of the quadratic base change, $U(1)$ symmetries in F-theory compactifications on K3 surfaces obtained as the quadratic base change of all rational elliptic surfaces times a K3 surface can be exhaustively studied. These investigations can be future directions. To show that the technique of the quadratic base change of rational elliptic surfaces to yield elliptic K3 surfaces can be useful in studying the structures of $U(1)$ gauge symmetries in F-theory compactifications, in this note, we particularly considered the rational elliptic surfaces with the singularity types (\ref{singularity rank 7 in 2})-(\ref{singularity rank 4 in 2}). The complex structure moduli of these particular rational elliptic surfaces have lower dimensions. As a result, these rational elliptic surfaces are described by the Weierstrass equations with the fewer numbers of parameters. Owing to this, relatively simple Weierstrass equations can describe these rational elliptic surfaces, and the resulting elliptic K3 surfaces obtained as the quadratic base change of these rational elliptic surfaces. The methods used in this study can be applied to other general rational elliptic surfaces.

\section{Rational elliptic surfaces with various Mordell--Weil ranks}
\label{sec3}

We construct families of rational elliptic surfaces with a global section, with various fixed singularity types. Each of these families is described by the Weierstrass equation with parameters. We consider the following singularity types: $E_7, D_7, E_6A_1, E_6, D_6, D_5, A_4$. 

\subsection{Rational elliptic surfaces with singularity types of rank 7}
\label{ssec3.1}
We construct rational elliptic surfaces with the singularity types $E_7, D_7$, and $E_6A_1$. The complex structure moduli of such rational elliptic surfaces are one-dimensional. As discussed in Section \ref{sssec4.2.1}, generic K3 surfaces obtained as the quadratic base change of these rational elliptic surfaces have the Mordell--Weil rank 1. 

\subsubsection{Singularity type $E_7$}
We construct a family of rational elliptic surfaces with a section with the singularity type $E_7$, a generic member of which has one type $III^*$ fiber and three type $I_1$ fibers \footnote{The configuration of the singular fibers, one type $III^*$ fiber, one type $II$ fiber, and one type $I_1$ fiber, is also possible for rational elliptic surfaces with $E_7$ singularity. The moduli of rational elliptic surfaces with a section with the singular fibers of this configuration is zero-dimensional. We do not consider the rational elliptic surfaces with this configuration of the singular fibers in this study.}. By considering the automorphism of the base $\P^1$, we may assume that the type $III^*$ fiber is located at the infinity of the base $\P^1$, and one of the three type $I_1$ fibers is at the origin of the base $\P^1$. The following Weierstrass equation \footnote{Both elliptic K3 surfaces and rational elliptic surfaces are elliptic fibrations over the base $\P^1$. Thus, the Weierstrass equations that we consider in this study can be seen as elliptic curves over the function field $\C(v)$, where $v:=t/s$.} describes this family:
\begin{equation}
\label{Weierstrass E7 in 3.1}
y^2=x^3+ (t-3s)\, s^3 x+ (a\, t-2s)\, s^5.
\end{equation}
We used $[t:s]$ to denote the homogeneous coordinate of the base $\P^1$. $a$ in the equation (\ref{Weierstrass E7 in 3.1}) is a parameter \footnote{Solving the condition that the Weierstrass equation of rational elliptic surfaces with a global section possesses one type $III^*$ fiber and three type $I_1$ fibers yields the solution with two parameters. As discussed in \cite{KimuraMizoguchi}, the coefficients of the Weierstrass equation $y^2=x^3+a_4 x+a_6$ rescale as:
\begin{eqnarray}
\label{rescale coeff in 3.1}
a_4 & \rightarrow & \beta^4\, a_4 \\ \nonumber
a_6 & \rightarrow & \beta^6\, a_6
\end{eqnarray}
under the rescaling:
\begin{eqnarray}
\label{rescale variable in 3.1}
x & \rightarrow & \frac{x}{\beta^2} \\ \nonumber
y & \rightarrow & \frac{y}{\beta^3}.
\end{eqnarray}
Thus, one of the two parameters are in fact redundant, and the two parameters can be reduced to one parameter under the rescaling (\ref{rescale coeff in 3.1}) and (\ref{rescale variable in 3.1}). Similar arguments as that stated here apply to rational elliptic surfaces with other types of singularities.}. The discriminant of the Weierstrass equation (\ref{Weierstrass E7 in 3.1}) is given as follows:
\begin{equation}
\label{disc E7 in 3.1}
\Delta=ts^9 \, [4t^2+(27a^2-36)st+(108-108a)s^2].
\end{equation}
We confirm from equations (\ref{Weierstrass E7 in 3.1}) and (\ref{disc E7 in 3.1}) that generic members of rational elliptic surfaces (\ref{Weierstrass E7 in 3.1}) have one type $III^*$ fiber and three type $I_1$ fibers. We show the correspondence of the vanishing orders of the coefficients $a_4, a_6$ of the Weierstrass equation $y^2=x^3+a_4 x+a_6$ and the types of the singular fibers in Table \ref{tabfiberandcoeff} below.

\begingroup
\renewcommand{\arraystretch}{1.5}
\begin{table}[htb]
\begin{center}
  \begin{tabular}{|c|c|c|c|} \hline
$
\begin{array}{c}
\mbox{Type of} \\
\mbox{singular fiber}
\end{array}
$
 & ord($a_4$) & ord($a_6$) & ord($\Delta$) \\ \hline
$I_0 $ & $\ge 0$ & $\ge 0$ & 0 \\ \hline
$I_n $  ($n\ge 1$) & 0 & 0 & $n$ \\ \hline
$II $ & $\ge 1$ & 1 & 2 \\ \hline
$III $ & 1 & $\ge 2$ & 3 \\ \hline
$IV $ & $\ge 2$ & 2 & 4 \\ \hline
$I_0^*$ & $\ge 2$ & 3 & 6 \\ \cline{2-3}
 & 2 & $\ge 3$ &  \\ \hline
$I_m^*$  ($m \ge 1$) & 2 & 3 & $m+6$ \\ \hline
$IV^*$ & $\ge 3$ & 4 & 8 \\ \hline
$III^*$ & 3 & $\ge 5$ & 9 \\ \hline
$II^*$ & $\ge 4$ & 5 & 10 \\ \hline   
\end{tabular}
\caption{\label{tabfiberandcoeff}Vanishing orders of the coefficients, and the discriminant $\Delta$, of the Weierstrass equation $y^2=x^3+a_4 x+a_6$, and the corresponding types of the singular fibers.}
\end{center}
\end{table}  
\endgroup

\subsubsection{Singularity type $D_7$}
We construct moduli of rational elliptic surfaces with a global section with the singularity type $D_7$, generic members of which have one type $I^*_3$ fiber and three type $I_1$ fibers. We may assume that the type $I^*_3$ fiber is located at the origin of the base $\P^1$. The following Weierstrass form describes the complex structure moduli of such rational elliptic surfaces:
\begin{equation}
\label{Weierstrass D7 in 3.1}
y^2=x^3+t^2\, (t^2-3s^2)\, x + t^3\, (a\, t^3+t^2s-2s^3).
\end{equation}
$a$ in the equation (\ref{Weierstrass D7 in 3.1}) is a parameter. The discriminant of the Weierstrass equation (\ref{Weierstrass D7 in 3.1}) is given as follows:
\begin{equation}
\label{disc D7 in 3.1}
\Delta=t^9\, [(4+27a^2)t^3+54a\, t^2s-9\, ts^2-108a\, s^3].
\end{equation}
We confirm from equations (\ref{Weierstrass D7 in 3.1}) and (\ref{disc D7 in 3.1}) that a generic member of the moduli (\ref{Weierstrass D7 in 3.1}) has one type $I^*_3$ fiber and three type $I_1$ fibers.

\subsubsection{Singularity type $E_6A_1$}
We construct the moduli of rational elliptic surfaces with a section with the singularity type $E_6A_1$, generic members of which have one type $IV^*$ fiber, one type $I_2$ fiber and two type $I_1$ fibers \footnote{The configuration of the singular fibers, one type $IV^*$ fiber, one type $III$ fiber, and one type $I_1$ fiber, is also possible for rational elliptic surfaces with the singularity type $E_6A_1$. The moduli of the rational elliptic surfaces with this configuration of the singular fibers is zero-dimensional. We do not consider the rational elliptic surfaces with this configuration of the singular fibers in this study.}. We may assume that the type $IV^*$ fiber is located at the infinity of the base $\P^1$. We may also assume that the type $I_2$ fibers is at the origin. The following Weierstrass equation describes this family:
\begin{equation}
\label{Weierstrass E6A1 in 3.1}
y^2=x^3 + (t-3s)\, s^3 x+ (at^2+ts-2s^2)\, s^4.
\end{equation}
$a$ in the equation (\ref{Weierstrass E6A1 in 3.1}) is a parameter. The discriminant of the Weierstrass equation (\ref{Weierstrass E6A1 in 3.1}) is given by
\begin{equation}
\label{disc E6A1 in 3.1}
\Delta=t^2s^8\, [27a^2\, t^2 +(54a+4)\, ts-(108a+9)s^2].
\end{equation}
We confirm from the equations (\ref{Weierstrass E6A1 in 3.1}) and (\ref{disc E6A1 in 3.1}) that a generic member of the moduli (\ref{Weierstrass E6A1 in 3.1}) has one type $IV^*$ fiber, one type $I_2$ fiber and two type $I_1$ fibers.

\subsection{Rational elliptic surfaces with singularity types of rank 6}
\label{ssec3.2}
We discuss families of rational elliptic surfaces with the singularity types $E_6$ and $D_6$. The complex structure moduli of these rational elliptic surfaces are two-dimensional. These rational elliptic surfaces have the Mordell--Weil rank 2. Generic K3 surfaces obtained as the quadratic base changes of these rational elliptic surfaces have the Mordell--Weil rank 2, as described in Section \ref{sssec4.2.2}.

\subsubsection{Singularity type $E_6$}
\label{sssec3.2.1}
We construct a family of rational elliptic surfaces, generic members of which have one type $IV^*$ fiber and four type $I_1$ fibers. This yields a family of rational elliptic surfaces with the singularity type $E_6$. Using the automorphism of the base $\P^1$, we may assume that the type $IV^*$ fiber is located at the infinity, and one of the four type $I_1$ fibers is located at the origin. The Weierstrass equation of such a family is given by the following equation:
\begin{equation}
\label{Weierstrass E6 in 3.2}
y^2=x^3+ s^3(a\, t-3s)\, x+ s^4(b\, t^2+c\, ts-2s^2).
\end{equation}
$a,b,c$ in the equation (\ref{Weierstrass E6 in 3.2}) are parameters. The discriminant of the Weierstrass form (\ref{Weierstrass E6 in 3.2}) is given by
\begin{equation}
\label{disc E6 in 3.2}
\Delta=ts^8 \, [27b^2\, t^3+(4a^3+54bc)\, t^2s+(-36a^2-108b+27c^2)\, ts^2+108(a-c)\, s^3].
\end{equation}
We confirm from the equations (\ref{Weierstrass E6 in 3.2}) and (\ref{disc E6 in 3.2}) that a generic member of the moduli (\ref{Weierstrass E6 in 3.2}) has one type $IV^*$ fiber and four $I_1$ fibers. 
\par We may also assume that one of the remaining three type $I_1$ fibers is at $[t:s]=[1:1]$ by the automorphism of the base $\P^1$. This assumption imposes the following constraint on the parameters $a,b,c$:
\begin{equation}
\label{condition E6 in 3.2}
4a^3-36a^2+108a+27\, [b^2+2b(c-2)+c(c-4)]=0.
\end{equation}
Therefore, the complex structure of the Weierstrass equation (\ref{Weierstrass E6 in 3.2}) is in fact determined by two parameters, and the remaining one parameter is redundant. The remaining one parameter is determined by the other two parameters and the constraint (\ref{condition E6 in 3.2}).

\subsubsection{Singularity type $D_6$}
We construct a family of rational elliptic surfaces, generic members of which have one type $I_2^*$ fiber and four type $I_1$ fibers. This yields a family of rational elliptic surfaces with the singularity type $D_6$. Using the automorphism of the base $\P^1$, we may assume that the type $I^*_2$ fiber is at the origin, and one of four type $I_1$ fibers is at the infinity. The following Weierstrass equation describes this family:
\begin{equation}
\label{Weierstrass D6 in 3.2}
y^2=x^3+ t^2(-t^2+a\, ts-3s^2)\, x+t^3(\frac{2}{3\sqrt{3}}\, t^3+b\, t^2s+a\, ts^2-2s^3).
\end{equation}
$a,b$ in the equation (\ref{Weierstrass D6 in 3.2}) are parameters. The discriminant of the Weierstrass equation (\ref{Weierstrass D6 in 3.2}) is given by
\begin{equation}
\label{disc D6 in 3.2}
\begin{split}
\Delta= & t^8s\cdot [ (-9a^2-108b-108)\, s^3+(4a^3+54ab+72a-24\sqrt{3})\, ts^2+ \\
& (-12a^2+12\sqrt{3}\, a+27b^2-36)\, t^2s+(12a+12\sqrt{3}\, b)\, t^3].
\end{split}
\end{equation}
We confirm from the equations (\ref{Weierstrass D6 in 3.2}) and (\ref{disc D6 in 3.2}) that a generic member of the moduli (\ref{Weierstrass D6 in 3.2}) has one type $I_2^*$ fiber and four $I_1$ fibers.

\subsection{Rational elliptic surfaces with singularity type of rank 5}
\label{ssec3.3}
We discuss families of rational elliptic surfaces with the singularity types $D_5$. The complex structure moduli of such rational elliptic surfaces is three-dimensional. These rational elliptic surfaces have the Mordell--Weil rank 3. Generic K3 surfaces obtained as the quadratic base changes of these rational elliptic surfaces have the Mordell--Weil rank 3, as discussed in Section \ref{sssec4.2.3}.
\par We construct the moduli of rational elliptic surfaces with a section with the singularity type $D_5$, generic members of which have one type $I^*_1$ fiber and five type $I_1$ fibers. Using the automorphism of the base $\P^1$, we may assume that the type $I^*_1$ fiber is at the origin, and one of the five type $I_1$ is at the infinity. The following Weierstrass equation describes this family:
\begin{equation}
\label{Weierstrass D5 in 3.3}
y^2=x^3+t^2(-t^2+a\, ts-3s^2)\, x+t^3(\frac{2}{3\sqrt{3}}\, t^3+b\, t^2s+c\, ts^2-2s^3).
\end{equation}
$a,b,c$ in the equation (\ref{Weierstrass D5 in 3.3}) are parameters. The discriminant of the Weierstrass equation (\ref{Weierstrass D5 in 3.3}) is given by
\begin{equation}
\label{disc D5 in 3.3}
\begin{split}
\Delta= & t^7s\cdot [ (108a-108c)\, s^4+(-36a^2-108b+27c^2-108)\, ts^3+ \\
& (4a^3+72a+54bc-24\sqrt{3})\, t^2s^2+(-12a^2+27b^2+12\sqrt{3}\, c-36)\, t^3s \\
& +(12a+12\sqrt{3}\, b)\, t^4].
\end{split}
\end{equation}
We confirm from the equations (\ref{Weierstrass D5 in 3.3}) and (\ref{disc D5 in 3.3}) that a generic member of the moduli (\ref{Weierstrass D5 in 3.3}) has one type $I_1^*$ fiber and five $I_1$ fibers.

\subsection{Rational elliptic surfaces with singularity type of rank 4}
\label{ssec3.4}
We discuss families of rational elliptic surfaces with the singularity type $A_4$. The complex structure moduli of such rational elliptic surfaces is four-dimensional. These rational elliptic surfaces have the Mordell--Weil rank 4. Generic K3 surfaces obtained as the quadratic base changes of these rational elliptic surfaces have the Mordell--Weil rank 4, as described in Section \ref{sssec4.2.4}.
\par We construct the moduli of rational elliptic surfaces with a section with the singularity type $A_4$, generic members of which have one type $I_5$ fiber and seven type $I_1$ fibers. Using the automorphism of the base $\P^1$, we may assume that the type $I_5$ fiber is located at the origin. This family is described by the following Weierstrass equation:
\begin{equation}
\label{Weierstrass A4 in 3.4}
\begin{split}
y^2= & \, x^3+(a\, t^4+b\, t^3s+c\, t^2s^2+d\, ts^3-3s^4)\, x+ \\
&  [e\, t^6+ f\, t^5s+ (a-\frac{c^2}{12}-\frac{bd}{6}-\frac{cd^2}{72}-\frac{d^4}{1728})t^4s^2+ \\
& (b-\frac{cd}{6}-\frac{d^3}{216})t^3s^3+(c-\frac{d^2}{12})t^2s^4+d\, ts^5-2s^6].
\end{split}
\end{equation}
$a,b,c,d,e,f$ in the equation (\ref{Weierstrass A4 in 3.4}) are parameters. The discriminant of the Weierstrass equation (\ref{Weierstrass A4 in 3.4}) is given by
\begin{equation}
\label{disc A4 in 3.4}
\begin{split}
\Delta= &  t^5 \cdot [(-18bc-18ad-\frac{3}{2}c^2d-\frac{3}{2}bd^2-\frac{1}{4}cd^3-\frac{d^5}{96}-108f)\, s^7+ \\
&  (-9b^2-18ac-\frac{c^3}{2}+6bcd+\frac{15}{2}ad^2+\frac{3}{8}c^2d^2+\frac{1}{2}bd^3+\frac{7}{96}cd^4+\frac{11d^6}{3456}-108e+54df)\, ts^6 + \\
& +(-18ab+\frac{15}{2}bc^2+3b^2d+15acd+\frac{3}{4}c^3d+\frac{3}{4}bcd^2-\frac{1}{4}ad^3+ \\
& \frac{7}{48}c^2d^3+\frac{bd^4}{96}+\frac{5cd^5}{576}+\frac{d^7}{6912}+54de+54cf-\frac{9}{2}d^2f)\, t^2s^5 + \\
& (-9a^2+12b^2c+\frac{15}{2}ac^2+\frac{3}{16}c^4+15abd+\frac{3}{4}bc^2d+\frac{3}{4}b^2d^2-\frac{3}{4}acd^2+ \\
& \frac{1}{16}c^3d^2+\frac{1}{8}bcd^3-\frac{1}{32}ad^4+\frac{c^2d^4}{128}+\frac{bd^5}{192}+\frac{cd^6}{2304}+\frac{d^8}{110592}+ \\
& 54ce-\frac{9}{2}d^2e+54bf-9cdf-\frac{d^3f}{4})\, t^3s^4 + \\
& (4b^3+24abc+12a^2d+54be-9cde-\frac{1}{4}d^3e+54af-\frac{9}{2}c^2f-9bdf-\frac{3}{4}cd^2f-\frac{d^4f}{32})\, t^4s^3 + \\
& (12ab^2+12a^2c+54ae-\frac{9}{2}c^2e-9bde-\frac{3}{4}cd^2e-\frac{d^4e}{32}+27f^2)\, t^5s^2 + \\
& (12a^2b+54ef)\, t^6s + (4a^3+27e^2)\, t^7 ].
\end{split}
\end{equation}
We confirm from the equations (\ref{Weierstrass A4 in 3.4}) and (\ref{disc A4 in 3.4}) that a generic member of the moduli (\ref{Weierstrass A4 in 3.4}) has one type $I_5$ fiber and seven $I_1$ fibers.
\par Using the automorphism of the base $\P^1$, we may assume that one of the seven type $I_1$ fibers is located at the infinity. This imposes the following constraint on the parameters:
\begin{equation}
4a^3+27e^2=0.
\end{equation}
Furthermore, we may fix another type $I_1$ fiber. We may, for example, require that one of the remaining six type $I_1$ fibers is located at $[t:s]=[1:1]$. Therefore, the actual parameters are four among the six parameters $a,b,c,d,e,f$. The remaining two are redundant, and they are determined by the four parameters. Therefore, the moduli (\ref{Weierstrass A4 in 3.4}) is parameterized by four parameters.

\section{Constructions of K3 surfaces with various Mordell-Weil ranks as quadratic base change of rational elliptic surfaces}
\label{sec4}
\subsection{Review of quadratic base change of rational elliptic surfaces}
We review the quadratic base change of the rational elliptic surfaces. We consider a rational elliptic surface with a global section, described by the following Weierstrass equation:
\begin{equation}
\label{rational elliptic surface in 4.1}
y^2=x^3+f\, x+g,
\end{equation}
where $f,g$ are homogeneous polynomials of the base $\P^1$ of degree 4 and 6, respectively. We denote the homogeneous coordinate of the base $\P^1$ by $[t:s]$. 
\par A quadratic base change is an operation that replaces $t,s$ with homogeneous quadratic polynomials:
\begin{eqnarray}
\label{quadratic base change in 4.1}
t & \rightarrow & \alpha_1 t^2+ \alpha_2 ts + \alpha_3 s^2 \\ \nonumber
s & \rightarrow & \alpha_4 t^2+ \alpha_5 ts + \alpha_6 s^2.
\end{eqnarray}
$\alpha_i$, $i=1, \cdots, 6$ are parameters. Homogeneous polynomials $f,g$ transform accordingly under the quadratic base change. We denote the resulting polynomials after the quadratic base change by $\til{f}, \til{g}$, respectively. Therefore, $\til{f}, \til{g}$ are homogeneous polynomials of degree 8 and 12, respectively. The following Weierstrass equation
\begin{equation}
\label{base change surface in 4.1}
y^2=x^3+\til{f}\, x+\til{g}
\end{equation}
describes an elliptic surface obtained as the quadratic base change of the rational elliptic surface (\ref{rational elliptic surface in 4.1}). As discussed in \cite{KRES}, the surface (\ref{base change surface in 4.1}) gives an elliptic K3 surface. The generic quadratic base change corresponds to gluing of two identical rational elliptic surfaces, and the singular fibers of the resulting K3 surface are twice as many as those of the original rational elliptic surface \cite{KRES}.
\par Considering the quadratic base change of the rational elliptic surfaces with nonzero Mordell--Weil ranks, which we obtained as described in Section \ref{sec3}, we construct families of K3 surfaces with nonzero Mordell--Weil ranks in Section \ref{ssec4.2}. We deduce the Weierstrass equations of the K3 surfaces obtained as the quadratic base change of rational elliptic surfaces in Section \ref{ssec4.2}. The Mordell--Weil ranks remain the same after the base change for generic parameters of the quadratic base change; the Mordell--Weil rank increases for special values of the parameters $\alpha_i$, $i=1, \cdots, 6$. Thus, we obtain families of K3 surfaces whose Mordell--Weil ranks are 1, 2, 3, and 4 for generic values of the parameters $\alpha_i$, $i=1, \cdots, 6$. 

\subsection{K3 surfaces with various Mordell--Weil ranks obtained as quadratic base change}
\label{ssec4.2}
We consider the quadratic base change of rational elliptic surfaces obtained as described in Section \ref{sec3} to construct families of elliptic K3 surfaces with various Mordell--Weil ranks. 

\subsubsection{K3 surfaces with Mordell--Weil rank 1}
\label{sssec4.2.1}
We constructed rational elliptic surfaces with the singularity types $E_7$, $D_7$, and $E_6 A_1$ as described in Section \ref{ssec3.1}. These rational elliptic surfaces have Mordell--Weil groups of rank 1. The quadratic base change of these rational elliptic surfaces yields families of elliptic K3 surfaces with the Mordell--Weil rank 1, the singularity types of which are twice those of the original rational elliptic surfaces; the resulting K3 surfaces generically \footnote{The singularity type enhances, or the Mordell--Weil rank increases at the special points in the moduli. These situations are discussed for complex structure moduli of elliptic K3 surfaces with a global section with the singularity types of rank 17 in \cite{KimuraMizoguchi}.} have the Mordell--Weil rank 1, with the singularity types $E^2_7$, $D^2_7$, and $E^2_6 A^2_1$. 
\par To be explicit, we particularly describe the Weierstrass equation of K3 surfaces with the singularity type $E^2_7$ with the Mordell--Weil rank 1, obtained as the quadratic base change of the family of rational elliptic surfaces (\ref{Weierstrass E7 in 3.1}) with the singularity type $E_7$. As discussed in Section \ref{ssec3.1}, the Weierstrass equation of rational elliptic surfaces with the singularity type $E_7$ is given as follows:
\begin{equation}
\label{RES E7 in 4.2.1}
y^2=x^3+ (t-3s)\, s^3 x+ (a\, t-2s)\, s^5.
\end{equation}
The variables $t,s$ transform as follows under the quadratic base change:
\begin{eqnarray}
\label{quadratic base change in 4.2.1}
t & \rightarrow & \alpha_1 t^2+ \alpha_2 ts + \alpha_3 s^2 \\ \nonumber
s & \rightarrow & \alpha_4 t^2+ \alpha_5 ts + \alpha_6 s^2.
\end{eqnarray}
Thus, the resulting K3 surfaces are given by the following Weierstrass equation:
\begin{equation}
\label{K3 E7 times 2 in 4.2}
\begin{split}
y^2= & x^3+ [(\alpha_1 t^2+ \alpha_2 ts + \alpha_3 s^2)-3(\alpha_4 t^2+ \alpha_5 ts + \alpha_6 s^2)]\, (\alpha_4 t^2+ \alpha_5 ts + \alpha_6 s^2)^3 x + \\
 &  [a\, (\alpha_1 t^2+ \alpha_2 ts + \alpha_3 s^2)-2(\alpha_4 t^2+ \alpha_5 ts + \alpha_6 s^2)]\, (\alpha_4 t^2+ \alpha_5 ts + \alpha_6 s^2)^5.
 \end{split}
\end{equation}
The elliptic K3 surface (\ref{K3 E7 times 2 in 4.2}) has two type $III^*$ fibers and six type $I_1$ fibers for generic values of the parameters $a$ and $\alpha_i$, $i=1,\cdots, 6$.
\par After some computations, we find that the rational elliptic surface with the singularity type $E_7$ (\ref{RES E7 in 4.2.1}) admits the following free section:
\begin{equation}
\label{free section of RES E7 in 4.2.1}
[X:Y:Z]=[-a s^2: \sqrt{3a-a^3-2}\cdot s^3 : 1].
\end{equation}
By comparing the general formula of a generating section of an elliptic surface with the Mordell--Weil group of rank 1 obtained in \cite{MorrisonPark} with the section (\ref{free section of RES E7 in 4.2.1}), we deduce that the section (\ref{free section of RES E7 in 4.2.1}) is in fact a generating section of the Mordell--Weil group of the rational elliptic surface with the singularity type $E_7$ (\ref{RES E7 in 4.2.1}). Two identical copies of the section (\ref{free section of RES E7 in 4.2.1}) are glued together to yield a section of the elliptic K3 surface (\ref{K3 E7 times 2 in 4.2}) that is obtained as the quadratic base change of the rational elliptic surface with the singularity type $E_7$ (\ref{RES E7 in 4.2.1}). Replacing $t,s$ in the section (\ref{free section of RES E7 in 4.2.1}) with homogeneous quadratic polynomials as in (\ref{quadratic base change in 4.2.1}), we obtain the expression for the resulting section of the elliptic K3 surface (\ref{K3 E7 times 2 in 4.2}) as follows:
\begin{equation}
\label{resulting section K3 in 4.2.1}
[X:Y:Z]=[-a\, (\alpha_4 t^2+ \alpha_5 ts + \alpha_6 s^2)^2: \sqrt{3a-a^3-2}\, (\alpha_4 t^2+ \alpha_5 ts + \alpha_6 s^2)^3 : 1].
\end{equation}
Because the quadratic base change (\ref{quadratic base change in 4.2.1}) induces an injective group homomorphism from the Mordell--Weil group of the rational elliptic surface (\ref{RES E7 in 4.2.1}) to the Mordell--Weil group of the elliptic K3 surface (\ref{K3 E7 times 2 in 4.2}), the resulting section (\ref{resulting section K3 in 4.2.1}) generates the group $\Z$.
\par The Weierstrass equations of the K3 surfaces with the singularity types $D_7^2$, $E^2_6A^2_1$, with the Mordell--Weil rank 1 can be obtained in ways similar to that described previously. The resulting elliptic K3 surfaces with the singularity type $D_7^2$, obtained as the quadratic base change of rational elliptic surfaces (\ref{Weierstrass D7 in 3.1}), are described by the following Weierstrass equation:
\begin{equation}
\label{K3 D7 times 2 in 4.2}
\begin{split}
y^2= & x^3+(\alpha_1 t^2+ \alpha_2 ts + \alpha_3 s^2)^2\cdot [(\alpha_1 t^2+ \alpha_2 ts + \alpha_3 s^2)^2-3(\alpha_4 t^2+ \alpha_5 ts + \alpha_6 s^2)^2]\, x + \\
&  (\alpha_1 t^2+ \alpha_2 ts + \alpha_3 s^2)^3\cdot [a\, (\alpha_1 t^2+ \alpha_2 ts + \alpha_3 s^2)^3 \\
& +(\alpha_1 t^2+ \alpha_2 ts + \alpha_3 s^2)^2(\alpha_4 t^2+ \alpha_5 ts + \alpha_6 s^2) -2(\alpha_4 t^2+ \alpha_5 ts + \alpha_6 s^2)^3].
\end{split}
\end{equation}
Elliptic K3 surface (\ref{K3 D7 times 2 in 4.2}) has two type $I^*_3$ fibers and six type $I_1$ fibers for generic values of the parameters $a$ and $\alpha_i$, $i=1,\cdots, 6$.  
\par Elliptic K3 surfaces with the singularity type $E_6^2 A_1^2$, obtained as the quadratic base change of rational elliptic surfaces (\ref{Weierstrass E6A1 in 3.1}), are described by the following Weierstrass equation:
\begin{equation}
\label{K3 E6A1 times 2 in 4.2}
\begin{split}
y^2= & x^3 + [(\alpha_1 t^2+ \alpha_2 ts + \alpha_3 s^2)-3(\alpha_4 t^2+ \alpha_5 ts + \alpha_6 s^2)]\, (\alpha_4 t^2+ \alpha_5 ts + \alpha_6 s^2)^3\, x+ \\
&  [a(\alpha_1 t^2+ \alpha_2 ts + \alpha_3 s^2)^2+(\alpha_1 t^2+ \alpha_2 ts + \alpha_3 s^2)(\alpha_4 t^2+ \alpha_5 ts + \alpha_6 s^2) \\
& -2(\alpha_4 t^2+ \alpha_5 ts + \alpha_6 s^2)^2]\cdot (\alpha_4 t^2+ \alpha_5 ts + \alpha_6 s^2)^4.
\end{split}
\end{equation}
Elliptic K3 surface (\ref{K3 E6A1 times 2 in 4.2}) has two type $IV^*$ fibers, two type $I_2$ fibers and four type $I_1$ fibers for generic values of the parameters $a$ and $\alpha_i$, $i=1,\cdots, 6$.

\subsubsection{K3 surfaces with Mordell--Weil rank 2}
\label{sssec4.2.2}
We constructed families of rational elliptic surfaces with the singularity types $E_6$ and $D_6$ in Section \ref{ssec3.2}. Quadratic base changes of these families yield families of elliptic K3 surfaces with the singularity types $E_6^2$ and $D_6^2$, respectively, the Mordell--Weil groups of which generically have the ranks 2. 
\par As described in Section \ref{sssec3.2.1}, family of rational elliptic surfaces with the singularity type $E_6$ is given by the following Weierstrass equation:
\begin{equation}
y^2=x^3+ s^3(a\, t-3s)\, x+ s^4(b\, t^2+c\, ts-2s^2).
\end{equation}
Applying the quadratic base change (\ref{quadratic base change in 4.1}), we deduce that family of elliptic K3 surfaces with the singularity type $E_6^2$ is described by the following Weierstrass equation:
\begin{equation}
\label{K3 E6 times 2 in 4.2}
\begin{split}
y^2= & x^3+ (\alpha_4 t^2+ \alpha_5 ts + \alpha_6 s^2)^3[a\, (\alpha_1 t^2+ \alpha_2 ts + \alpha_3 s^2)-3(\alpha_4 t^2+ \alpha_5 ts + \alpha_6 s^2)]\, x+ \\
& (\alpha_4 t^2+ \alpha_5 ts + \alpha_6 s^2)^4 \\
& \cdot [b\, (\alpha_1 t^2+ \alpha_2 ts + \alpha_3 s^2)^2+c\, (\alpha_1 t^2+ \alpha_2 ts + \alpha_3 s^2)(\alpha_4 t^2+ \alpha_5 ts + \alpha_6 s^2)\\
& -2(\alpha_4 t^2+ \alpha_5 ts + \alpha_6 s^2)^2].
\end{split}
\end{equation}
Elliptic K3 surface (\ref{K3 E6 times 2 in 4.2}) has two type $IV^*$ fibers and eight type $I_1$ fibers for generic values of the parameters $a,b$ \footnote{As discussed in Section \ref{sssec3.2.1}, $c$ is determined by the parameters $a$ and $b$.} and $\alpha_i$, $i=1,\cdots, 6$.
\par The Weierstrass equation of family of elliptic K3 surfaces with the singularity type $D_6^2$ can be obtained by considering the quadratic base change (\ref{quadratic base change in 4.1}) of rational elliptic surfaces with the singularity type $D_6$ (\ref{Weierstrass D6 in 3.2}). The Weierstrass equation of elliptic K3 surfaces with the singularity type $D_6^2$ obtained as the quadratic base change of rational elliptic surfaces (\ref{Weierstrass D6 in 3.2}) is given as follows:
\begin{equation}
\label{K3 D6 times 2 in 4.2}
\begin{split}
y^2= & x^3+ (\alpha_1 t^2+ \alpha_2 ts + \alpha_3 s^2)^2\cdot [-(\alpha_1 t^2+ \alpha_2 ts + \alpha_3 s^2)^2+ \\
& a\, (\alpha_1 t^2+ \alpha_2 ts + \alpha_3 s^2)(\alpha_4 t^2+ \alpha_5 ts + \alpha_6 s^2) -3(\alpha_4 t^2+ \alpha_5 ts + \alpha_6 s^2)^2]\, x+ \\
& (\alpha_1 t^2+ \alpha_2 ts + \alpha_3 s^2)^3\cdot [\frac{2}{3\sqrt{3}}\, (\alpha_1 t^2+ \alpha_2 ts + \alpha_3 s^2)^3+ \\
& b\, (\alpha_1 t^2+ \alpha_2 ts + \alpha_3 s^2)^2(\alpha_4 t^2+ \alpha_5 ts + \alpha_6 s^2)+ \\
& a\, (\alpha_1 t^2+ \alpha_2 ts + \alpha_3 s^2)(\alpha_4 t^2+ \alpha_5 ts + \alpha_6 s^2)^2-2(\alpha_4 t^2+ \alpha_5 ts + \alpha_6 s^2)^3].
\end{split}
\end{equation}
Elliptic K3 surface (\ref{K3 D6 times 2 in 4.2}) has two type $I^*_2$ fibers and eight type $I_1$ fibers for generic values of the parameters $a,b$ and $\alpha_i$, $i=1,\cdots, 6$.

\subsubsection{K3 surfaces with Mordell--Weil rank 3}
\label{sssec4.2.3}
The quadratic base change of rational elliptic surfaces with the singularity type $D_5$ that we constructed in Section \ref{ssec3.3} yields a family of elliptic K3 surfaces with the singularity type $D_5^2$, the Mordell--Weil groups of which generically have the rank 3. 
\par As described in Section \ref{ssec3.3}, the Weierstrass equation of family of rational elliptic surfaces with the singularity type $D_5$ is given by
\begin{equation}
\label{Weierstrass D5 in 4.2.3}
y^2=x^3+t^2(-t^2+a\, ts-3s^2)\, x+t^3(\frac{2}{3\sqrt{3}}\, t^3+b\, t^2s+c\, ts^2-2s^3).
\end{equation}
The quadratic base change (\ref{quadratic base change in 4.1}) of the equation (\ref{Weierstrass D5 in 4.2.3}) of rational elliptic surfaces with the singularity type $D_5$ yields the Weierstrass equation of the resulting K3 surfaces with the singularity type $D_5^2$ as follows:
\begin{equation}
\label{K3 D5 times 2 in 4.2.3}
\begin{split}
y^2= & x^3+ \\
& (\alpha_1 t^2+ \alpha_2 ts + \alpha_3 s^2)^2\cdot [-(\alpha_1 t^2+ \alpha_2 ts + \alpha_3 s^2)^2 \\
& +a\, (\alpha_1 t^2+ \alpha_2 ts + \alpha_3 s^2)(\alpha_4 t^2+ \alpha_5 ts + \alpha_6 s^2) -3(\alpha_4 t^2+ \alpha_5 ts + \alpha_6 s^2)^2]\, x+ \\
& (\alpha_1 t^2+ \alpha_2 ts + \alpha_3 s^2)^3\cdot [\frac{2}{3\sqrt{3}}\, (\alpha_1 t^2+ \alpha_2 ts + \alpha_3 s^2)^3 \\
& +b\, (\alpha_1 t^2+ \alpha_2 ts + \alpha_3 s^2)^2(\alpha_4 t^2+ \alpha_5 ts + \alpha_6 s^2)+ \\
& c\, (\alpha_1 t^2+ \alpha_2 ts + \alpha_3 s^2)(\alpha_4 t^2+ \alpha_5 ts + \alpha_6 s^2)^2-2(\alpha_4 t^2+ \alpha_5 ts + \alpha_6 s^2)^3].
\end{split}
\end{equation}
Elliptic K3 surface (\ref{K3 D5 times 2 in 4.2.3}) has two type $I^*_1$ fibers and ten type $I_1$ fibers for generic values of the parameters $a,b,c$ and $\alpha_i$, $i=1,\cdots, 6$.

\subsubsection{K3 surfaces with Mordell--Weil rank 4}
\label{sssec4.2.4}
Family of rational elliptic surfaces with the singularity type $A_4$ was constructed in Section \ref{ssec3.4}. Quadratic base change of this family yields a family of elliptic K3 surfaces with the singularity type $A_4^2$, the Mordell--Weil groups of which generically have the rank 4. 
\par As described in Section \ref{ssec3.4}, the Weierstrass equation of family of rational elliptic surfaces with the singularity type $A_4$ is given by
\begin{equation}
\label{Weierstrass A4 in 4.2}
\begin{split}
y^2= & \, x^3+(a\, t^4+b\, t^3s+c\, t^2s^2+d\, ts^3-3s^4)\, x+ \\
&  [e\, t^6+ f\, t^5s+ (a-\frac{c^2}{12}-\frac{bd}{6}-\frac{cd^2}{72}-\frac{d^4}{1728})t^4s^2+ \\
& (b-\frac{cd}{6}-\frac{d^3}{216})t^3s^3+(c-\frac{d^2}{12})t^2s^4+d\, ts^5-2s^6].
\end{split}
\end{equation}
The quadratic base change (\ref{quadratic base change in 4.1}) of rational elliptic surfaces (\ref{Weierstrass A4 in 4.2}) with the singularity type $A_4$ yields the following Weierstrass equation of the resulting K3 surfaces with the singularity type $A_4^2$:
\begin{equation}
\label{K3 A4 times 2 in 4.2}
\begin{split}
y^2= & \, x^3+ \\
& [a\, (\alpha_1 t^2+ \alpha_2 ts + \alpha_3 s^2)^4+b\, (\alpha_1 t^2+ \alpha_2 ts + \alpha_3 s^2)^3(\alpha_4 t^2+ \alpha_5 ts + \alpha_6 s^2)+ \\
& c\, (\alpha_1 t^2+ \alpha_2 ts + \alpha_3 s^2)^2(\alpha_4 t^2+ \alpha_5 ts + \alpha_6 s^2)^2+ \\
& d\, (\alpha_1 t^2+ \alpha_2 ts + \alpha_3 s^2)(\alpha_4 t^2+ \alpha_5 ts + \alpha_6 s^2)^3-3(\alpha_4 t^2+ \alpha_5 ts + \alpha_6 s^2)^4]\cdot x+ \\
&  [e\, (\alpha_1 t^2+ \alpha_2 ts + \alpha_3 s^2)^6+ f\, (\alpha_1 t^2+ \alpha_2 ts + \alpha_3 s^2)^5(\alpha_4 t^2+ \alpha_5 ts + \alpha_6 s^2)+ \\
&  (a-\frac{c^2}{12}-\frac{bd}{6}-\frac{cd^2}{72}-\frac{d^4}{1728})(\alpha_1 t^2+ \alpha_2 ts + \alpha_3 s^2)^4(\alpha_4 t^2+ \alpha_5 ts + \alpha_6 s^2)^2+ \\
& (b-\frac{cd}{6}-\frac{d^3}{216})(\alpha_1 t^2+ \alpha_2 ts + \alpha_3 s^2)^3(\alpha_4 t^2+ \alpha_5 ts + \alpha_6 s^2)^3+ \\
& (c-\frac{d^2}{12})(\alpha_1 t^2+ \alpha_2 ts + \alpha_3 s^2)^2(\alpha_4 t^2+ \alpha_5 ts + \alpha_6 s^2)^4+ \\
& d\, (\alpha_1 t^2+ \alpha_2 ts + \alpha_3 s^2)(\alpha_4 t^2+ \alpha_5 ts + \alpha_6 s^2)^5-2(\alpha_4 t^2+ \alpha_5 ts + \alpha_6 s^2)^6].
\end{split}
\end{equation}
Elliptic K3 surface (\ref{K3 A4 times 2 in 4.2}) has two type $I_5$ fibers and fourteen type $I_1$ fibers for generic values of the parameters $a,b,c,d$ \footnote{As discussed in Section \ref{ssec3.4}, $e,f$ are determined by the parameters $a,b,c$ and $d$.} and $\alpha_i$, $i=1,\cdots, 6$.

\section{F-theory compactifications on the constructed families of K3 surfaces}
\label{sec5}
In this section, we discuss deducing the structures of the gauge symmetries that arise in F-theory compactifications on the families of K3 surfaces with nonzero Mordell--Weil ranks, constructed as described in Section \ref{ssec4.2}, times a K3 surface. Anomaly cancellation conditions are discussed in Section \ref{ssec5.2}.

\subsection{Global structures of gauge symmetries in F-theory compactifications}

\subsubsection{K3 surfaces with Mordell--Weil rank 1}
Generic members of the family of K3 surfaces described by equation (\ref{K3 E7 times 2 in 4.2}) have two type $III^*$ fibers and six type $I_1$ fibers with the Mordell--Weil rank 1. Therefore, an $E_7^2 \times U(1)$ gauge group arises in F-theory compactifications on a generic K3 surface of the family (\ref{K3 E7 times 2 in 4.2}) times a K3. Torsion parts of the Mordell--Weil groups of the elliptic K3 surfaces with a global section are classified in \cite{Shimada}. The torsion part of the Mordell--Weil group of an elliptic K3 surface with the singularity type $E_7^2$ is trivial (No. 856 in Table 1 in \cite{Shimada}). Therefore, the Mordell--Weil group of a generic member of the family (\ref{K3 E7 times 2 in 4.2}) is isomorphic to $\Z$, and the global structure of the gauge group in F-theory compactification is 
\begin{equation}
E^2_7 \times U(1).
\end{equation}

\vspace{1cm}

\par Generic K3 surfaces of the family (\ref{K3 D7 times 2 in 4.2}) have the singularity type $D^2_7$ with the Mordell--Weil rank 1. The torsion part of an elliptic K3 surface with the singularity type $D_7^2$ is 0 \cite{Shimada}; the Mordell--Weil group of a generic member of the family (\ref{K3 D7 times 2 in 4.2}) is isomorphic to $\Z$. By a similar argument as that stated previously, we deduced that the global structure of the gauge symmetry that arises in F-theory compactifications on generic K3 surfaces of the family (\ref{K3 D7 times 2 in 4.2}) times a K3 surface is 
\begin{equation}
SO(14)^2 \times U(1).
\end{equation}

\vspace{1cm}

\par Generic members of the family (\ref{K3 E6A1 times 2 in 4.2}) have the singularity type $E_6^2 A_1^2$ with Mordell--Weil rank 1. The torsion part of the Mordell--Weil group of an elliptic K3 surface with singularity type $E^2_6 A^2_1$ is trivial \cite{Shimada}; therefore, the Mordell--Weil group of a generic K3 surface of the family (\ref{K3 E6A1 times 2 in 4.2}) is isomorphic to $\Z$. Thus, we find that the global structure of the gauge group that arises in F-theory compactifications on generic K3 surfaces of the family (\ref{K3 E6A1 times 2 in 4.2}) times a K3 surface is 
\begin{equation}
E^2_6 \times SU(2)^2 \times U(1).
\end{equation}

\subsubsection{K3 surfaces with Mordell--Weil rank 2}
Generic K3 surfaces of the family (\ref{K3 E6 times 2 in 4.2}) have the singularity type $E_6^2$ with Mordell--Weil rank 2, as described in Section \ref{sssec4.2.2}. The torsion part of the Mordell--Weil group of a generic K3 surface (\ref{K3 E6 times 2 in 4.2}) can be determined from the singularity type $E_6^2$; it is 0 \cite{Shimada}. Thus, the Mordell--Weil group of a generic elliptic K3 surface (\ref{K3 E6 times 2 in 4.2}) is isomorphic to $\Z^2$. We deduce from these results that the global structure of the gauge group that arises in F-theory compactifications on generic elliptic K3 surfaces (\ref{K3 E6 times 2 in 4.2}) times a K3 surface is 
\begin{equation}
E^2_6 \times U(1)^2.
\end{equation}

\vspace{1cm}

\par Generic K3 surfaces of the family (\ref{K3 D6 times 2 in 4.2}) have the singularity type $D_6^2$ with Mordell--Weil rank 2. The torsion part of the Mordell--Weil group  of an elliptic K3 surface with the singularity type $D_6^2$ is trivial according to Table 1 in \cite{Shimada}. Therefore, the Mordell--Weil group of a generic elliptic K3 surface (\ref{K3 D6 times 2 in 4.2}) is isomorphic to $\Z^2$. Thus, we find that the global structure of the gauge group that arises in F-theory compactifications on generic elliptic K3 surfaces (\ref{K3 D6 times 2 in 4.2}) times a K3 is 
\begin{equation}
SO(12)^2 \times U(1)^2.
\end{equation}

\subsubsection{K3 surfaces with Mordell--Weil rank 3}
Generic K3 surfaces (\ref{K3 D5 times 2 in 4.2.3}) have the singularity type $D_5^2$ with the Mordell--Weil rank 3. Thus, it follows from Table 1 in \cite{Shimada} that the torsion parts of their Mordell--Weil groups are 0. Thus, the Mordell--Weil groups of generic K3 surfaces (\ref{K3 D5 times 2 in 4.2.3}) are isomorphic to $\Z^3$. We deduce from these that the global structure of the gauge group that arises in F-theory compactifications on generic elliptic K3 surfaces (\ref{K3 D5 times 2 in 4.2.3}) times a K3 is 
\begin{equation}
SO(10)^2 \times U(1)^3.
\end{equation}

\subsubsection{K3 surfaces with Mordell--Weil rank 4}
Generic K3 surfaces of the family (\ref{K3 A4 times 2 in 4.2}) have the singularity type $A_4^2$ with the Mordell--Weil rank 4. It follows that the torsion parts of their Mordell--Weil groups are trivial \cite{Shimada}; therefore, the Mordell--Weil groups of generic K3 surfaces of the family (\ref{K3 A4 times 2 in 4.2}) are isomorphic to $\Z^4$. Thus, we conclude that the global structure of the gauge group that arises in F-theory compactifications on generic elliptic K3 surfaces (\ref{K3 A4 times 2 in 4.2}) times a K3 is 
\begin{equation}
SU(5)^2 \times U(1)^4.
\end{equation}

\subsection{Cancellation of anomaly in F-theory compactifications}
\label{ssec5.2}
F-theory compactified on elliptically fibered spaces constructed as the direct products of K3 surfaces without a four-form flux yields four-dimensional theory with $N=2$ supersymmetry. Cancellation of the tadpole \cite{SVW} without a flux determines the form of the discriminant locus of the compactification space, constructed as the direct product of K3 surfaces, to be the sum of 24 K3 surfaces. 7-branes are wrapped on these K3 surfaces; there are 24 7-branes in F-theory compactification on the direct product of K3 surfaces \footnote{See, for example, \cite{K} for discussion of F-theory compactifications on direct products of K3 surfaces, and the form of the discriminant locus of the direct product of K3 surfaces.}. 
\par The Euler number of a singular fiber gives the number of 7-branes associated with that singular fiber. The Euler numbers of the types of the singular fibers of an elliptic fibration can be found in \cite{Kod2}. The Euler numbers of the types of the singular fibers of an elliptic fibration are presented in Table \ref{fiber7-brane}.

\begingroup
\renewcommand{\arraystretch}{1.1}
\begin{table}[htb]
\centering
  \begin{tabular}{|c|c|} \hline
Singular fiber & $\begin{array}{c}
\mbox{\# of 7-branes} \\
\mbox{(Euler number)}
\end{array}$ \\ \hline
$I_n$ & $n$\\
$I^*_0$ &  6\\ 
$I^*_m$ &  6$+m$\\ 
$II $ &  2\\
$III $ & 3\\
$IV $ & 4\\
$IV^*$ & 8 \\ 
$III^*$ & 9 \\
$II^*$ & 10 \\ \hline
\end{tabular}
\caption{\label{fiber7-brane}Associated numbers of 7-branes for types of singular fibers.}
\end{table}
\endgroup

\par We confirm that the anomaly cancellation condition is satisfied for F-theory compactifications on the K3 surfaces constructed in Section \ref{sec4} times a K3 surface. 
\par We consider F-theory compactification on the K3 surface (\ref{K3 E7 times 2 in 4.2}) times a K3. The compactification space has two type $III^*$ fibers and six type $I_1$ fibers. Thus, we confirm from Table \ref{fiber7-brane} that the number of 7-branes associated with the singular fibers is 24 in total. This shows that the anomaly cancellation condition is in fact satisfied for F-theory compactification on the K3 surface (\ref{K3 E7 times 2 in 4.2}) times a K3.
\par By similar arguments as that stated here, it follows that the anomaly cancellation condition is satisfied for F-theory compactifications on the remaining K3 surfaces constructed as described in Section \ref{ssec4.2} times a K3. 

\section{Conclusions}
\label{sec6}
In this study, we constructed families of rational elliptic surfaces with a global section, which have the Mordell--Weil ranks 1, 2, 3, and 4, utilizing the fact that the sum of the rank of the singularity type and the rank of the Mordell--Weil group of any rational elliptic surface with a section is 8. By considering the gluing of pairs of identical rational elliptic surfaces, we obtained families of elliptic K3 surfaces, the ranks of the Mordell--Weil groups of which are 1, 2, 3, and 4. We determined the Weierstrass equations to describe these families of elliptic K3 surfaces. We deduced the structures of the gauge symmetries that arise in F-theory compactifications on the obtained K3 surfaces times a K3 surface. $U(1)$ gauge fields arise in these F-theory compactifications. Promoting the results obtained in this note concerning elliptic K3 surfaces to elliptic Calabi--Yau 3-folds and elliptic Calabi--Yau 4-folds can be future directions.

\section*{Acknowledgments}

We would like to thank Shun'ya Mizoguchi and Shigeru Mukai for discussions. This work is partially supported by Grant-in-Aid for Scientific Research {\#}16K05337 from the Ministry of Education, Culture, Sports, Science and Technology of Japan.


\begin{thebibliography}{99}

\bibitem{Vaf}C.~Vafa, ``Evidence for F-theory'', {\it Nucl. Phys.} {\bf B 469} (1996) 403 [arXiv:hep-th/9602022].
\bibitem{MV1}D.~R.~Morrison and C.~Vafa, ``Compactifications of F-theory on Calabi-Yau threefolds. 1'', {\it Nucl. Phys.} {\bf B 473} (1996) 74 [arXiv:hep-th/9602114].
\bibitem{MV2}D.~R.~Morrison and C.~Vafa, ``Compactifications of F-theory on Calabi-Yau threefolds. 2'', {\it Nucl. Phys.} {\bf B 476} (1996) 437 [arXiv:hep-th/9603161].
\bibitem{Witten}E.~Witten, ``Strong coupling expansion of Calabi-Yau compactification'', {\it Nucl. Phys.} {\bf B471} (1996) 135--158 [arXiv:hep-th/9602070].
\bibitem{DWmodel}R.~Donagi and M.~Wijnholt, ``Model Building with F-Theory'', {\it Adv. Theor. Math. Phys.} {\bf 15} (2011) no.5, 1237--1317 [arXiv:0802.2969 [hep-th]].
\bibitem{BHV1}C.~Beasley, J.~J.~Heckman and C.~Vafa, ``GUTs and Exceptional Branes in  F-theory -I'', {\it JHEP} {\bf 01} (2009) 058 [arXiv:0802.3391 [hep-th]].
\bibitem{BHV2}C.~Beasley, J.~J.~Heckman and C.~Vafa, ``GUTs and Exceptional Branes in F-theory - II: Experimental Predictions'', {\it JHEP} {\bf 01} (2009) 059 [arXiv:0806.0102 [hep-th]].
\bibitem{DWGUT}R.~Donagi and M.~Wijnholt, ``Breaking GUT Groups in F-Theory'', {\it Adv. Theor. Math. Phys.} {\bf 15} (2011) 1523--1603 [arXiv:0808.2223 [hep-th]].

\bibitem{MorrisonPark}D.~R.~Morrison and D.~S.~Park, ``F-Theory and the Mordell-Weil Group of Elliptically-Fibered Calabi-Yau Threefolds'', {\it JHEP} {\bf 10} (2012) 128 [arXiv:1208.2695 [hep-th]].
\bibitem{MPW}C.~Mayrhofer, E.~Palti and T.~Weigand, ``U(1) symmetries in F-theory GUTs with multiple sections'', {\it JHEP} {\bf 03} (2013) 098 [arXiv:1211.6742 [hep-th]].
\bibitem{BGK}V.~Braun, T.~W.~Grimm and J.~Keitel, ``New Global F-theory GUTs with U(1) symmetries'', {\it JHEP} {\bf 09} (2013) 154 [arXiv:1302.1854 [hep-th]].
\bibitem{BMPWsection}J.~Borchmann, C.~Mayrhofer, E.~Palti and T.~Weigand, ``Elliptic fibrations for $SU(5)\times U(1)\times U(1)$ F-theory vacua'', {\it Phys. Rev.} {\bf D88} (2013) no.4 046005 [arXiv:1303.5054 [hep-th]].
\bibitem{CKP}M.~Cveti\v c, D.~Klevers and H.~Piragua, ``F-Theory Compactifications with Multiple U(1)-Factors: Constructing Elliptic Fibrations with Rational Sections'', {\it JHEP} {\bf 06} (2013) 067 [arXiv:1303.6970 [hep-th]].
\bibitem{BGK1306}V.~Braun, T.~W.~Grimm and J.~Keitel, ``Geometric Engineering in Toric F-Theory and GUTs with U(1) Gauge Factors,'' {\it JHEP} {\bf 12} (2013) 069 [arXiv:1306.0577 [hep-th]].
\bibitem{CGKP}M.~Cveti\v c, A.~Grassi, D.~Klevers and H.~Piragua, ``Chiral Four-Dimensional F-Theory Compactifications With SU(5) and Multiple U(1)-Factors'', {\it JHEP} {\bf 04} (2014) 010 [arXiv:1306.3987 [hep-th]].
\bibitem{CKPS}M.~Cveti\v c, D.~Klevers, H.~Piragua and P.~Song, ``Elliptic fibrations with rank three Mordell-Weil group: F-theory with U(1) x U(1) x U(1) gauge symmetry,'' {\it JHEP} {\bf 1403} (2014) 021 [arXiv:1310.0463 [hep-th]].
\bibitem{AL}I.~Antoniadis and G.~K.~Leontaris, ``F-GUTs with Mordell-Weil U(1)'s,'' {\it Phys. Lett.} {\bf B735} (2014) 226--230 [arXiv:1404.6720 [hep-th]].
\bibitem{LSW}C.~Lawrie, S.~Sch\"afer-Nameki and J.-M.~Wong, ``F-theory and All Things Rational: Surveying U(1) Symmetries with Rational Sections'', {\it JHEP} {\bf 09} (2015) 144 [arXiv:1504.05593 [hep-th]]. 
\bibitem{CKPT}M.~Cveti\v c, D.~Klevers, H.~Piragua and W.~Taylor, ``General U(1)$\times$U(1) F-theory compactifications and beyond: geometry of unHiggsings and novel matter structure,'' {\it JHEP} {\bf 1511} (2015) 204 [arXiv:1507.05954 [hep-th]].
\bibitem{CGKPS}M.~Cveti\v c, A.~Grassi, D.~Klevers, M.~Poretschkin and P.~Song, ``Origin of Abelian Gauge Symmetries in Heterotic/F-theory Duality,'' {\it JHEP} {\bf 1604} (2016) 041 [arXiv:1511.08208 [hep-th]].
\bibitem{KimuraMizoguchi}Y.~Kimura and S.~Mizoguchi, ``Enhancements in F-theory models on moduli spaces of K3 surfaces with $ADE$ rank 17'', {\it PTEP} {\bf 2018} no. 4 (2018) 043B05 [arXiv:1712.08539 [hep-th]].

\bibitem{KRES}Y.~Kimura, ``Structure of stable degeneration of K3 surfaces into pairs of rational elliptic surfaces'', {\it JHEP} {\bf 03} (2018) 045 [arXiv:1710.04984 [hep-th]].
\bibitem{FMW}R.~Friedman, J.~Morgan and E.~Witten, ``Vector bundles and F theory'', {\it Commun. Math. Phys.} {\bf 187} (1997) 679--743 [arXiv:hep-th/9701162].
\bibitem{AM}P.~S.~Aspinwall and D.~R.~Morrison, ``Point - like instantons on K3 orbifolds'', {\it Nucl. Phys.} {\bf B503} (1997) 533--564 [arXiv:hep-th/9705104].
\bibitem{AHK}L.~B.~Anderson, J.~J.~Heckman and S.~Katz, ``T-Branes and Geometry'', {\it JHEP} {\bf 05} (2014) 080 [arXiv:1310.1931 [hep-th]].
\bibitem{BKW}A.~P.~Braun, Y.~Kimura and T.~Watari, ``The Noether-Lefschetz problem and gauge-group-resolved landscapes: F-theory on K3 $\times$ K3 as a test case'', {\it JHEP} {\bf 04} (2014) 050 [arXiv:1401.5908 [hep-th]].

\bibitem{DolgachevZhang}I.~V.~Dolgachev and D.-Q.~Zhang, ``Coble rational surfaces'', {\it Amer. Math. J.} {\bf 123} (2001) 79--114 [arXiv:math/9909135]. 
\bibitem{CantatDolgachev}S.~Cantat and I.~V.~Dolgachev, ``Rational surfaces with a large group of automorphisms'', {\it J. Amer. Math. Soc.} {\bf 25} (2012) 863--905 [arXiv:1106.0930 [math.AG]].

\bibitem{Kimura1801}Y.~Kimura, ``K3 surfaces without section as double covers of Halphen surfaces, and F-theory compactifications'', {\it PTEP} {\bf 2018} (2018) 043B06 [arXiv:1801.06525 [hep-th]].

\bibitem{OS}K.~Oguiso and T.~Shioda, ``The Mordell-Weil lattice of a rational elliptic surface'', {\it Comment. Math. Univ. St. Pauli} {\bf 40} (1991) 83.
\bibitem{MP}R.~Miranda and U.~Persson, ``On extremal rational elliptic surfaces'', {\it Math. Z.} {\bf 193} (1986) 537--558.
\bibitem{Nar}I.~Naruki, ``Configurations related to maximal rational elliptic surfaces'', {\it Adv. Stud. Pure Math.} {\bf 8} (1986) 315--347. 
\bibitem{Kurumadani}Y.~Kurumadani, ``Pencils of cubic curves and rational elliptic surfaces,'' {\it RIMS Preprints}, RIMS-1800 (2014). 

\bibitem{BIKMSV}M.~Bershadsky, K.~A.~Intriligator, S.~Kachru, D.~R.~Morrison, V.~Sadov and C.~Vafa, ``Geometric singularities and enhanced gauge symmetries'', {\it Nucl. Phys.} {\bf B 481} (1996) 215 [arXiv:hep-th/9605200].
\bibitem{MT}S.~Mizoguchi and T.~Tani, ``Looijenga's weighted projective space, Tate's algorithm and Mordell-Weil Lattice in F-theory and heterotic string theory'', {\it JHEP} {\bf 11} (2016) 053 [arXiv:1607.07280 [hep-th]].

\bibitem{Kod1}K.~Kodaira, ``On compact analytic surfaces II'', {\it Ann. of Math.} {\bf 77} (1963), 563--626.
\bibitem{Kod2}K.~Kodaira, ``On compact analytic surfaces III'', {\it Ann. of Math.} {\bf 78} (1963), 1--40.
\bibitem{Ner}A.~N\'eron, ``Mod\`eles minimaux des vari\'et\'es ab\'eliennes sur les corps locaux et globaux'', {\it Publications math{\' e}matiques de l'IH{\' E}S} {\bf 21} (1964), 5--125.
\bibitem{Tate}J.~Tate, ``Algorithm for determining the type of a singular fiber in an elliptic pencil'', in {\it Modular Functions of One Variable IV}, Springer, Berlin (1975), 33--52.
\bibitem{Shiodamodular}T.~Shioda, ``On elliptic modular surfaces'', {\it J. Math. Soc. Japan} {\bf 24} (1972), 20--59.
\bibitem{Shioda}T.~Shioda, ``On the Mordell-Weil lattices'', {\it Comment. Math. Univ. St. Pauli} {\bf 39} (1990), 211--240.  
\bibitem{Silv}J.~H.~Silverman, {\it Advanced Topics in the Arithmetic of Elliptic Curves}, Graduate Texts in Mathematics {\bf 151}, Springer (1994).
\bibitem{BHPV}W.~Barth, K.~Hulek, C.~Peters and A.~Van de Ven, {\it Compact complex surfaces}, second edition, Springer (2004).
\bibitem{SchShio}M.~Sch\"{u}tt and T.~Shioda, ``Elliptic Surfaces'', in {\it Algebraic Geometry in East Asia (Seoul 2008)}, {\it Advanced Studies in Pure Mathematics} {\bf 60} (2010) 51--160 [arXiv:0907.0298 [math.AG]].
\bibitem{Tate1}J.~Tate, ``Algebraic cycles and poles of zeta functions'', in {\it Arithmetical Algebraic Geometry (Proc. Conf. Purdue Univ., 1963)}, Harper \& Row (1965), 93--110.
\bibitem{Tate2}J.~Tate, ``On the conjectures of Birch and Swinnerton-Dyer and a geometric analog'', {\it S\'eminaire Bourbaki} {\bf 9} (1964--1966), Expos\'e no. 306, 415--440.

\bibitem{Shimada}I.~Shimada, ``On elliptic K3 surfaces'', {\it Michigan Math. J.} {\bf 47} (2000) 423--446 [arXiv:math/0505140 [math.AG]].

\bibitem{SVW}S.~Sethi, C.~Vafa and E.~Witten, ``Constraints on low dimensional string compactifications'', {\it Nucl. Phys.} {\bf B 480} (1996) 213--224, [arXiv: hep-th/9606122].
\bibitem{K}Y.~Kimura, ``Gauge Groups and Matter Fields on Some Models of F-theory without Section'', {\it JHEP} {\bf 03} (2016) 042 [arXiv:1511.06912 [hep-th]].

\end{thebibliography}
\end{document}